\documentclass[10pt]{bmc_article}

\usepackage{cite} 
\usepackage{url}  
\usepackage{ifthen}  
\usepackage{multicol}   
\usepackage[utf8]{inputenc} 
\usepackage[margin=0.8in]{geometry}
\usepackage{amsmath}  
\usepackage{amssymb}
\usepackage{color}    
\usepackage{graphicx} 
\usepackage{hyperref} 
\usepackage{versions}
\usepackage{rotating}

\usepackage{soul}

\newcommand{\xVec}[1]{\ensuremath{\mathbf{#1}}}
\excludeversion{stare}

\input{glossary}
\newboolean{publ}

\newenvironment{bmcformat}{\begin{raggedright}\baselineskip20pt\sloppy\setboolean{publ}{false}}{\end{raggedright}\baselineskip20pt\sloppy}

\begin{document}
\begin{bmcformat}
%
\title{Computational model of sphingolipids metabolism:\\ a case study of Alzheimer's disease}
 %

\author{Agata Charzy\'{n}ska$^{\spadesuit,1,2}$%
       \email{a.charzynska@phd.ipipan.waw.pl}%
      \and
         Weronika Wronowska$^{\spadesuit, 3}$%
         \email{ wwro@biol.uw.edu.pl}
          \and
         Karol Niena{\l}towski$^4$%
         \email{k.nienaltowski@sysbiosig.org}
          \and
         Anna Gambin\correspondingauthor$^{5,2}$%
         \email{aniag@mimuw.edu.pl}%
      }
      


\address{%
   \iid(\spadesuit) {\bf Equal contributors} \\
   \ \\
    \iid(1)Institute of Computer Science Polish Academy of Sciences, Warsaw, Poland;\\
    \iid(2)Bioinformatics Laboratory, Mossakowski Medical Research Centre Polish Academy of Sciences, Warsaw, Poland;\\
    \iid(3)Faculty of Biology University of Warsaw, Warsaw, Poland;\\
    \iid(4)Institute of Fundamental Technological Research Polish Academy of Sciences, Warsaw, Poland;\\
    \iid(5)Institute of Informatics, University of Warsaw, Warsaw, Poland;
    }
\maketitle

\begin{abstract}
              \paragraph*{Background:}  Sphingolipids -- as suggested by the prefix in their name -- are mysterious molecules, which play surprisingly various roles in opposable cellular processes, like autophagy, apoptosis, proliferation and differentiation. Recently they have been also recognized as important messengers in cellular signalling pathways. More importantly, sphingolipid metabolism disorders were observed in various  pathological conditions such as cancer and  neurodegeneration.

        \paragraph*{Results:} Existing formal models of sphingolipids metabolism concentrates mostly on {\em de novo} ceramide synthesis or restrict  their focus to  biochemical transformations of a particular subspecies. We propose first comprehensive computational model of sphingolipid metabolism in human tissue. In contrast to previous approaches we explicitly model compartmentalization what allows emphasizing the differences  among individual organelles.

        \paragraph*{Conclusions:}  Presented here  model was validated by means of recently proposed model analysis technics allowing for detection of most sensitive and experimentally non-identifiable parameters and determination of main sources of model variance.
Moreover, we demonstrate the utility of  the model for the study of molecular processes underlying Alzheimer's disease.
\end{abstract}

\ifthenelse{\boolean{publ}}{\begin{multicols}{2}}{}



\section{ Background}

Sphingolipids (\abbr{SL}) are categorized as a class of complex lipids containing sphingoid base (\abbr{SPH})~\cite{thudichum}. Modification of this basic structure by amide-linked fatty acid or by phosphorylation gives rise to vast family of bioactive sphingolipids like: ceramide (\abbr{CER}), ceramide-1-phosphate(\abbr{C1P}),  sphingosine-1-phosphate (\abbr{S1P}) or sphingomyelin (\abbr{SM})~\cite{carter, pruett}. 
Ceramide is known to be branching point for the metabolism of various sphingolipids subspecies. There are three major pathways of ceramide synthesis. In \textit{de novo} synthesis pathway ceramide is created from less complex molecules~\cite{denovo}. The second path is catabolism of complex sphingolipids, mainly sphingomyelin~\cite{smcat}. Ceramide generation can also occur through the breakdown of complex sphingolipids that are ultimately broken down into sphingosine in acidic environment of lysosome. In this pathway, known as {\em salvage pathway}, sphingosine is then reused by reacylation to form ceramide~\cite{salv}. On the same time ceramide may serve as a substrate for the synthesis of \abbr{SM}, \abbr{C1P}, and \abbr{SPH} which, in turns, can be phosphorylated to \abbr{S1P}~\cite{kolter, hanprinc, gault, bartke, merrill}.
For the long time these molecules were considered to have mainly structural function, but only in the last two decades sphingolipids were recognized as important messengers in cellular signalling pathways~\cite{signaling1, signaling2}. \\
Notable body of work has been devoted studying the influence of sphingolipids metabolism on cellular fate – autophagy, apoptosis, proliferation and differentiation~\cite{han96, han02}. 
Importantly, individual sphingolipid species appear to have the opposite effect on cell growth and survival. Dynamic balance between proapoptotic molecules (e.g. \abbr{CER} and \abbr{SPH}) and antiapoptotic -- prosurvival ones (e.g. \abbr{S1P} and \abbr{C1P}) is termed as \textit{sphingolipid rheostat}~\cite{rheostat}. Disruptions in metabolic pathways involved in controlling this balance are considered to underlay various diseases. Indeed, sphingolipids are known to have critical implications for the pathogenesis and treatment of such different illnesses like cancer~\cite{ogr, ponn, ryland, beckham} and neurodegenerative disorders (e.g. Alzheimer's disease)~\cite{al1, al2, al3, al4, al5}.

\paragraph*{Related research.}
Formal modelling seems to be an excellent tool to predict the response of the systems in different scenarios and to the wide range of both external and internal perturbations. However, due to the complexity of sphingolipids metabolom, and the paucity of data, not much has been done in the field of computational sphingolipidom modelling. There are only  few models of \abbr{SL} metabolism available in the literature. Provided by Vasquez et al.~\cite{vasquez} model refers to \textit{de novo} ceramide synthesis~in yeast. It captures all essential elements of the ceramide synthesis from nonsphingolipid metabolites. However, no further steps of ceramide and other more complex sphingolipids recycling (\abbr{SM} catabolic pathway and {\em salvage pathway}) were taken into consideration. \\
The model proposed be Gupta et al.~\cite{gupta} describes the C16-branch of sphingolipid metabolism in RAW264.7 cells. An advantage of this model is that it combines lipidomics and transcriptomics data provided by the LIPID MAPS Consortium. However, the model is restricted to de closest metabolites of C16 ceramide. 
Moreover none of the proposed models captured cell compartmental division, although it is known that the ceramide metabolism differentiates among different cell compartments such as mitochondria, nucleus and cell membrane.
 Therefore we found the construction of computational model for the metabolism of complex sphingolipids in human tissues to be an appealing task. 
 
\paragraph*{Our results.}
 We propose a formal model of regulatory processes capturing sphingolipid metabolism pathways. Computational modelling is based on ordinary differential equations (\abbr{ODEs}) describing the evolution of species concentration. Kinetics of the model is mostly based on the Mass Action Law (\abbr{MAL}) for the molecular transportation reactions and the Michaelis-Menten (\abbr{MM}) approach for enzymatically catalysed reactions. The modelled kinetics also covers the inhibition within competing species.

According to our best knowledge it is the first computational model of sphingolipid metabolism comprising the compartmentalization based on typical structure of non-differentiated eukaryotic cell. Reactions parameters were estimated basing on publicly available  literature data and some default assumptions based on experience with Biochemical Systems Theory (\abbr{BST})~\cite{bst}, while the initial concentrations of particular sphingolipd species in each organelle were taken from the LIPID MAPS database~\cite{lipidmaps}. 
To validate our model we applied both standard and novel methods of analysis i.e.: the local sensitivity analysis~\cite{chara}, the variance decomposition ~\cite{komor2013} and
the clustering of model's parameters based on sensitivities clusters~\cite{Wlodarczyk13}. 
Finally we demonstrate the utility of our model to study the molecular events that are known to underly Alzheimer's disease (\abbr{AD}).
The proposed model provides a comprehensive, functional integration of experimental data, and we hope it will have significant implications for understanding of  still not fully elucidated dependence between sphingolipid metabolism and various diseases.
Moreover for the first time the two recently published methods of computational models analysis (i.e. variance decomposition~\cite{komor2013} and sensitivity clustering~\cite{Wlodarczyk13}) were applied to the medium size realistic biochemical model.


\section{Results and Discussion}

\subsection{Model of sphingolipids metabolism.}
\begin{sidewaysfigure}
\begin{center}
\textbf{The ceramides metabolism diagram}
\end{center}
\includegraphics[width=\textwidth]{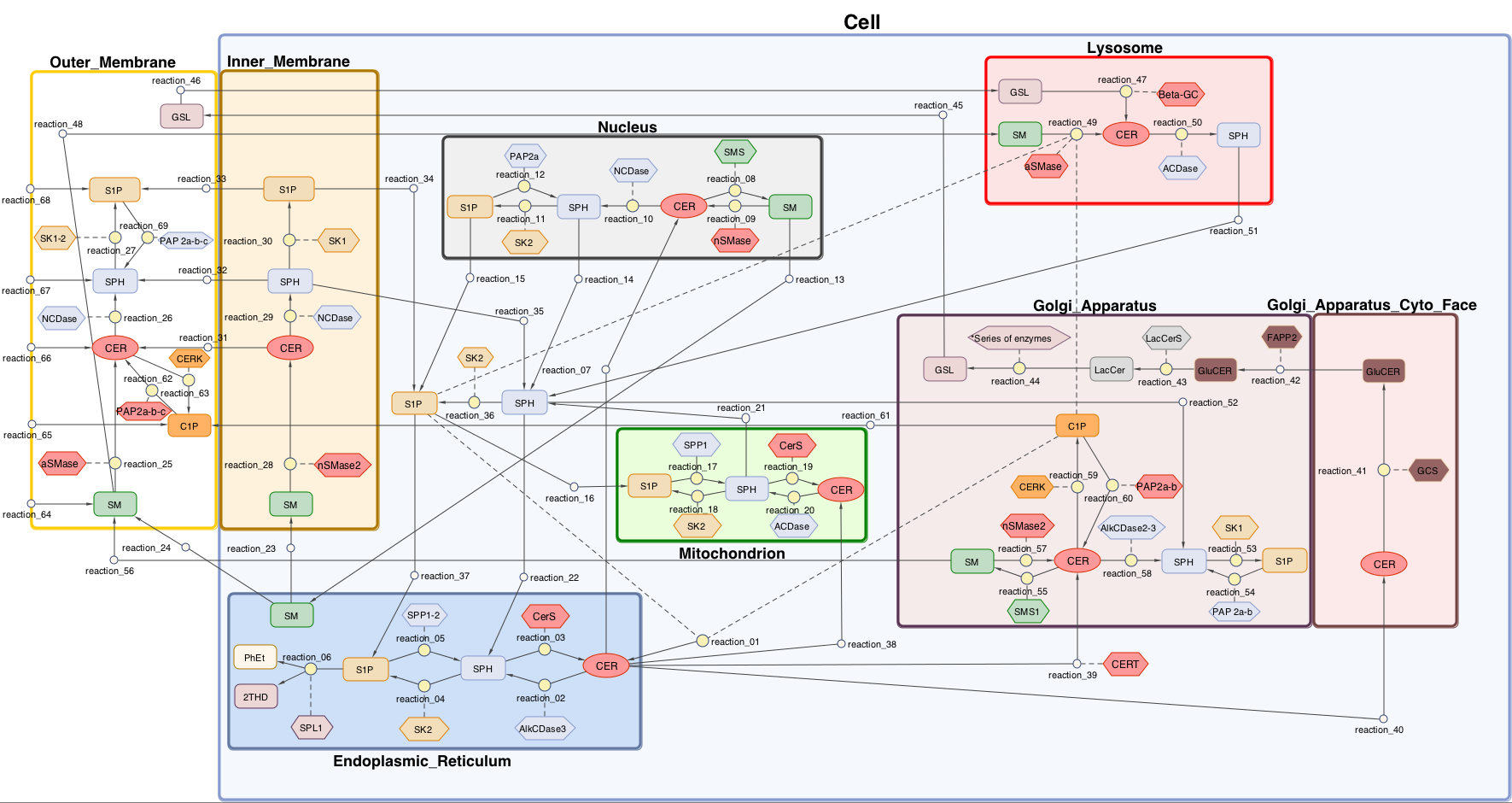}
\caption{Network of the SL metabolism system. The full model contains 69~reactions, 39~modelled species and 37~reaction catalysing enzymes. Solid lines indicate transportation and enzymatic reactions, dotted lines indicate inhibition.  }\label{fig:diagram}
\end{sidewaysfigure}

Our model captures all essential elements in the complex network of sphingolipids metabolism, except the \textit{de~novo} synthesis of ceramide, which had been described by Vasquez et al.~\cite{vasquez}. It illustrates the general behaviour of chosen subspecies in non specified human tissue in 9 sub-cellular compartments representing following organelle or their distinct parts:outer and inner layer of cell membrane, cytoplasm, endoplasmic reticulum, cytoplasmatic and lumenal face of Golgi apparatus, nucleus, mitochondria and lysosome. Our model includes 69 reactions of molecular transportation and biochemical transformation, see Figure~\ref{fig:diagram}.

\subsubsection*{Transportation}

 We applied Mass Action Law principle to describe transportation kinetics. Particular equations simulate different ways of \abbr{SL} transportation, which are determined by specific biophysical properties of particular sphingolipids~\cite{son}. It should be mentioned that most of these molecules (i.e. \abbr{CER}, \abbr{SM}, \abbr{GSL}) are restricted to biological membranes. Their transportation between organelle is possible only in form of complexes with lipid transfer proteins (\abbr{LTP})~\cite{lev} e.g. ceramide transfer protein (\abbr{CERT}) binds to \abbr{CER}
. Additionally, they might change their location in form of vesicles, as an integral part of biological membranes
~\cite{meer02}. For example translocation of \abbr{SM} and \abbr{GSL} from Golgi apparatus to outer membrane is correlated with exocytosis
, while during the endocytosis  complex \abbr{SL} may get in to the lysosome
. Sphingolipids may also diffuse along the membranes between linked organelle, as it is in case of ceramide floating between endoplasmic reticulum and nucleus
~\cite{egg}.  On the other hand, \abbr{SPH}, \abbr{S1P} and \abbr{C1P} are enough hydrophilic to diffuse freely from the membranes to the cytosol
, likewise from the outer membrane to the external environment
~\cite{tani}. Still, it was reported that transportation of \abbr{C1P} from the Golgi apparatus to other cellular compartments may also occur in association with specific transporter proteins as ceramide-1-phosphate transfer protein
 \abbr{CPTP}~\cite{sim}. Water solubility also determines molecules ability to flip between membrane leaflets. \abbr{CER} has relatively rapid flip rate
, also \abbr{SPH} is sufficiently amphipathic to move between membrane layers
~\cite{cont, meer11}. Finally \abbr{S1P} needs specific lipid transporters to traverse membranes
~\cite{kob, aye}. Similarly complex sphingolipids are unable to cross the membranes without the aid of specific flippases like four-phosphate adaptor protein 2 \abbr{FAPP2} puling \abbr{GluCER} from the outer to the inner surface of Golgi cisterns
~\cite{angel}.

\subsubsection*{Ceramide synthesis and degradation}

Most of the reactions depicted in the diagram are enzymatic. Michaelis-Menten model (\abbr{MM}) and simplified kinetics derived by the \abbr{MM} model was applied to describe different ways of synthesis and degradation of chosen \abbr{SL} species.  \abbr{CER} synthesis via \textit{de novo} path is described as the inflow of these molecule in endoplasmic reticulum
. \abbr{CER} might be also generated by the acetylation of \abbr{SPH}.  This reaction, catalysed by different types of ceramide synhases (\abbr{CerS})
~\cite{levy10} is the final step of the \textit{salvage pathway}
~\cite{salv}. Importantly  endoplasmic \abbr{SPH} utilized in this path may becomes from degradation of \abbr{S1P} what is catalysed by specific phospathases (\abbr{SPP1} and\abbr{SPP2} )
~\cite{brin04} or from lysosomal degradation of complex \abbr{SL}
. This later path initiated by acidic sphingomyelinase (\abbr{aSMase})
 and known to be critical for the maintaining proper concentrations of cellular \abbr{SL}~\cite{jenk09, kolt05}. In addition to described above endoplasmic route of \abbr{CER} synthesis similar subset of reactions can be described for the mitochondria. 
 Although, reactions of mitochondrial \abbr{SL} metabolism are not fully understood yet. Especially enzymes specificity and values of reaction rates parameters are often unknown~\cite{sisk05, bion04}. The third way of \abbr{CER} synthesis is hydrolysis of \abbr{SM}. \abbr{SMase}s responsible for catalysing this reaction are classified in to three categories based on their pH optima and sub-cellular distribution. Degradation of \abbr{SM} is known to be essential for the homeostasis of cell membranes, it was also reported to be strongly related to the stress induced apoptosis~\cite{sant96, pena97, han96}. Finally we described hydrolysis of \abbr{CER}.
Catalysing this reaction ceramidases \abbr{CDase}, 7 of which are known in human, cleave fatty acids from \abbr{CER} producing \abbr{SPH}~\cite{mao08}.

\subsubsection*{Synthesis of complex SL}

At the apex of \abbr{SL} complexity are sphingomyelins (\abbr{SM}) and even more diverse glycosphingolipids (\abbr{GSL}).  Even though some enzymes responsible for the synthesis of these complex \abbr{SL} were detected in e.g. nucleus 
 this pathway is mainly localized in the Golgi apparatus. In both cases \abbr{CER} is utilized as a backbone molecule. However, whether it would be converted in to \abbr{SM} or \abbr{GSL} depends on the way of its transportation from the endoplasmic reticulum. \abbr{CER} transported in the complex whit \abbr{CERT} protein gets to the cis-Golgi 
 where it gives rise to the \abbr{SM} in the \abbr{SMS} catalysed reaction
~\cite{merr90, tafe06}. While, to form \abbr{GSL} in the series reactions 
 \abbr{CER} must get to the trans-Golgi via vesicle dependent way
~\cite{fun}.  

\subsubsection*{\abbr{S1P} and \abbr{C1P} metabolism }

Moreover our model includes reactions of \abbr{CER} and \abbr{SPH} phosphorylation, produced \abbr{S1P} and \abbr{C1P}, unlike \abbr{CER} and \abbr{SPH}, promote cell growth and have anti-apoptotic properties~\cite{rheostat, cuvi96}. The antagonistic effect of these metabolites is regulated by the activity of many enzymes: (i) ceramide kinase \abbr{CERK} responsible for the synthesis of \abbr{C1P} 
 in the Golgi apparatus and the plasma membrane\cite{cerk}, (ii) sphingosine kinases \abbr{SK1} and \abbr{SK2} catalysing phosphorylation of \abbr{SPH} in different subcellular locations 
~\cite{mace02, pits11} (iii) mentioned before phosphatases able to hydrolyse \abbr{S1P} 
and \abbr{C1P}
. Among them are both lipid phosphate phosphatases of broad specificity (\abbr{PAP2a}, \abbr{PAP2b} and \abbr{PAP2c}) and \abbr{S1P} specific phosphatases (\abbr{SPP1} and \abbr{SPP2})~\cite{spie03, brin04}. All these enzymes with their isoformes differ in the substrate specificity, optimum pH and sub-cellular localization. Our model illustrates most of their known properties. For detailed characteristics see reviev articles~\cite{kolter, hanprinc, gault, bartke, merrill}. Importantly both \abbr{S1P} and \abbr{C1P} has been identified as inhibitors of enzymes responsible for \abbr{CER} synthesis like acidic sphingomyelinase (\abbr{aSMase}) and serine palmitoyltransferase (\abbr{SPT}), the key regulatory enzyme of \textit{de novo} synthesis pathway. This inhibitory activity was described in our model by the inhibitory kinetics, for details see supplementary Table~1.
Finally our model includes reaction of irreversible degradation of \abbr{S1P}
. Catalysed by sphingosine-1-phosphate lyase (\abbr{SPL1}) reaction of \abbr{S1P} hydrolysis to hexadecenal and phosphoetanolamine allows removal of sphingoid base from the pool of \abbr{SL} metabolites~\cite{bour10, spie03}.

 \subsubsection*{Model parameters}     

To conclude the model consists of 39 variables representing molecular species concentrations (some of them are the same compounds localized in different compartments see Figure \ref{fig:diagram}). Metabolic reactions network covers 69 biochemical enzymatic and transportation reactions among the reacting species. The model is implemented in a form of a system of 69 ordinary differential equations (\abbr{ODEs}) modelling the reaction network dynamics. Relevant input of the model are 129 parameters of inhibition and reactions rates in stationary state representing homeostasis conditions presented in supplementary Table~1 as well as 38 initial values of species concentrations (supplementary Table~2).  To achieve conditions resembling intracellular environment during the homoeostasis we stabilize the species concentrations to the stationary state of the system. Initial values for lipid levels were taken from the LIPID MAPS~\cite{vasquez}. In the sequel, we validate the model by performing local sensitivity analysis, variance decomposition and clustering analysis.

\subsection{Computational validation of the model}
 
 Biochemical models are characterized by substantially larger number of parameters relative to data size. Therefore the exact estimation of model parameters is very difficult to perform \cite{Brown03}. Thus we used mathematical modelling to analyse the dependencies between parameters and model dynamics.
To verify assumptions of our model we applied several techniques to get the broad view of the modelled system behaviour under normal and stressed conditions. The validation techniques were based on recently proposed and classical approaches engaging exact mathematical methods.

  \subsubsection*{Local sensitivity analysis}

The outcome of the local sensitivity analysis~\cite{chara}  performed for system in  stationary state -- homeostasis (see Figure \ref{fig:LSA_CER} and supplementary Figures 1-3) yielded the following conclusions.

\begin{center}
\textbf{The local sensitivity analysis of the ceramides species}
\begin{figure}[!ht]
\includegraphics[width=1.1\textwidth]{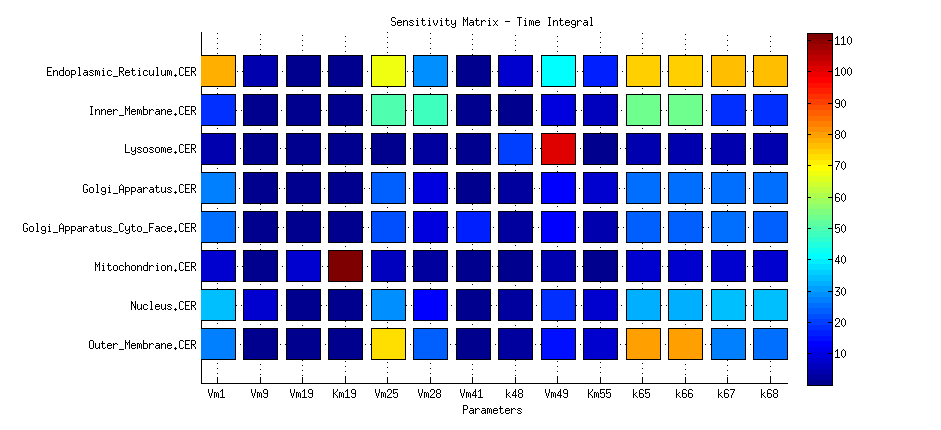}
\caption{The local sensitivity analysis of the CER species to the highly significant parameters.}
\label{fig:LSA_CER}
\end{figure}
\end{center}

\begin{itemize}
\item Within ceramide species the highest sensitivity indices were assigned to mitochondrial and lysosomal \abbr{CER} according to ceramide shynthase (\abbr{CerS}) in mitochondrion  and sphingomielynase (\abbr{SMase}) in lysosome respectively. The highest variability present \abbr{CER} in endoplasmic reticulum mostly to the model's inflows parameters both the exogenous by outer membrane and endogenous by \textit{de novo} synthesis in endoplasmic reticulum. What is more unexpected \abbr{CER} in endoplasmic reticulum is also highly sensitive to the parameter reaction catalysed by \abbr{SMase} in outer membrane. High sensitivity to the parameters of exogenous inflow of \abbr{CER} and \abbr{C1P} show membrane \abbr{CER} species, which are also sensitive to the membrane reactions catalysed by \abbr{SMase}s (see Figure  \ref{fig:LSA_CER}).
\item For the sphingosine species the greatest instability behaviour shows the mitochondrial \abbr{SPH}, that is highly sensitive to the model's inflows parameters as well as the parameter of reaction catalysed by enzyme \abbr{CerS} in mitochondrion. Moreover mitochondrial \abbr{SPH} is sensitive to the \abbr{SMase} catalysed reactions in membrane (see supplementary Figure~1).
In contrary concentration of \abbr{SPH} localised in cytozol  is practically invariant to parameters.
\item Mitochondrial \abbr{S1P} shows the greatest instability within the \abbr{S1P} species, not only for the model's inflows parameters and \abbr{CerS} in mitochondrion, but also for parameters of reactions catalysed by \abbr{SMase} in membrane and lysosom. For the other \abbr{S1P} species the greatest significant is sphingosinokinase in reticulum, inner membrane and nucleus respectively. Again the cytoplasmic \abbr{S1P} is the most stable \abbr{S1P} species (see supplementary Figure~2).
\item Sphingomyelin in outer membrane is the dominant species within all other species, consequently the exogenous inflow of \abbr{SM} by outer membrane is the most significant parameter for the \abbr{SM} species. Other surprising observation is that this parameter does not  play noticeable role for all other  species that are sensitive for model's inflows parameters by outer membrane (exogenous: \abbr{C1P}, \abbr{CER}, \abbr{SPH}, \abbr{S1P}). Another interesting feature is that the nuclear \abbr{SM} is the most stable within \abbr{SM} species (see supplementary Figure~3).

\end{itemize}

  \subsubsection*{Variance decomposition – homeostasis}

  The variance decomposition method \cite{komor2013} enables to decompose noise -- the uncertainty of the modelled output into components seaming from different reactions\cite{komor2013, jetka}. This method applied to our model  principally  indicate the reactions corresponding to edges in Figure~\ref{fig:diagram} incident to investigated species as the highest noise generators. Nevertheless some  reactions were more significant for investigated species than other incident reactions, whereas for some other species variance were distributed equally among all reactions. To find the distinctive reactions we calculated the mean variance for each investigated species and set the threshold to 110\% of the mean variance. The results for \abbr{CER} species  are  depicted in Figure~\ref{fig:SD_CER} and for \abbr{SPH}, \abbr{S1P} and \abbr{SM} is depicted in supplementary Figures~4-6, 
  respectively.

\begin{figure}[!ht]
\begin{center}
\textbf{The variance decomposition for the ceramides species}
\end{center} 
\includegraphics[width=\textwidth]{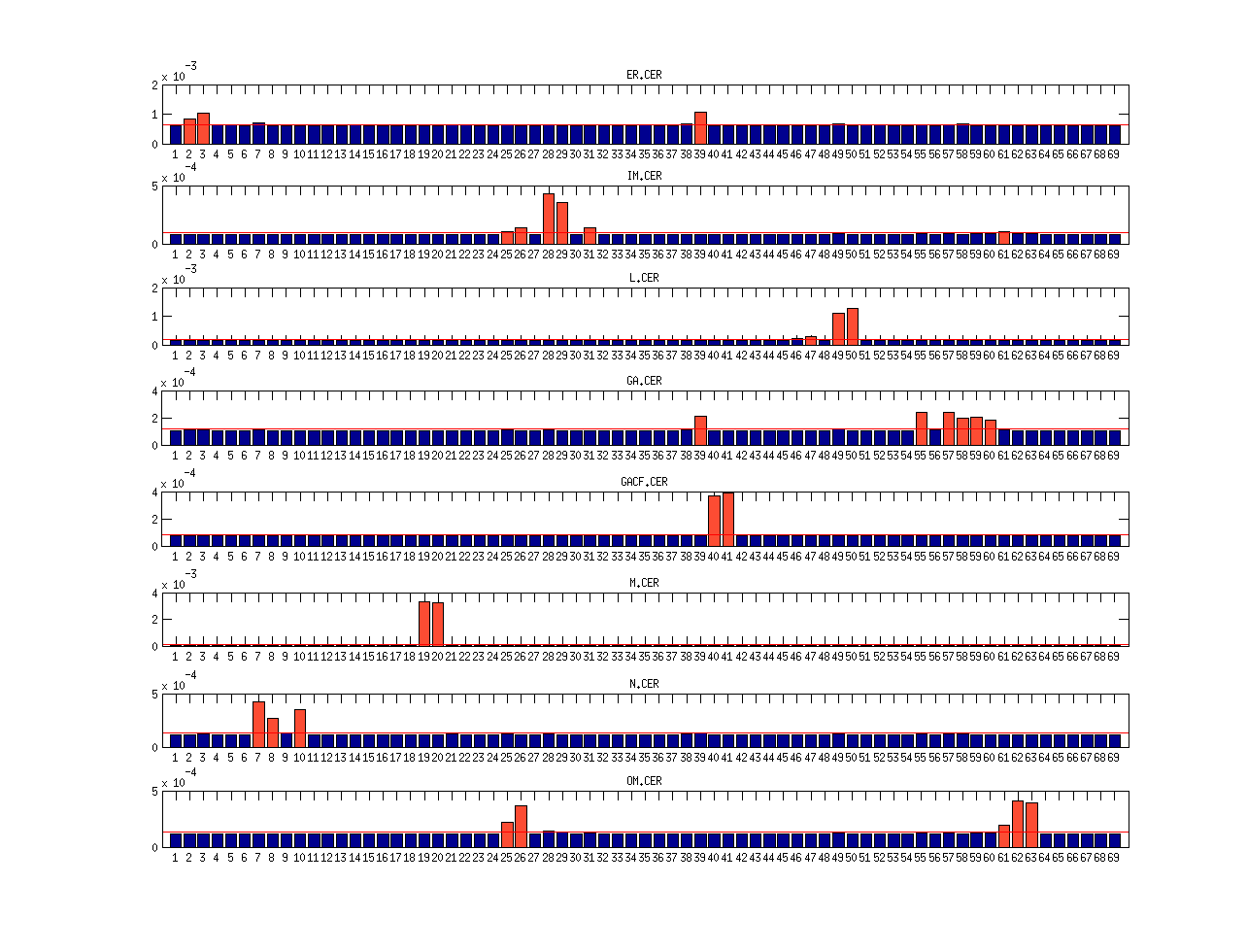}
\caption{ The variance decomposition of the ceramides concentration into components steaming from all model reactions. The red lines denotes the averages variance components of the investigated species. The red bars denotes the variance components that exceed the threshold of 110\% of average.  The x- axis denotes reactions numbers, the y-axis denotes the size of variance components.}\label{fig:SD_CER}
\end{figure}

\begin{itemize}
\item Within the ceramides species the highest variance shows the mitochondrial and lysosomal \abbr{CER}. For \abbr{CER} the threshold set on 110\% of average variance was exceeded only by the reactions 
catalysed by ceramideshynthase (\abbr{CerS}) and acid ceramidase (\abbr{ACDase}) in mitochondrion. 
The membrane \abbr{CER} species interact together so for inner membrane \abbr{CER} not only incident reactions 
exceeded the threshold but also reaction incident with outer membrane \abbr{CER} catalysed by \abbr{aSMase}. 
For the outer membrane \abbr{CER} the highest variance component steams from the reactions incident with outer membrane \abbr{C1P} and transportation reaction of \abbr{C1P} from Golgi apparatus to outer membrane. For the other \abbr{CER} species the highest variance is caused by the incident  reactions.
\item Within the sphingosine species the highest variance similarly to \abbr{CER} species falls to the mitochondrial \abbr{SPH}, whereas contrary to the mitochondrial \abbr{CER} for the mitochondrial \abbr{SPH} all reactions' noise components are near to the average. Interesting is that two incident reactions 
connecting mitochondrial \abbr{SPH} with mitochondrial \abbr{CER} exceeded threshold and two other incident reactions 
connecting mitochondrial \abbr{SPH} with mitochondrial \abbr{S1P} are significantly below the average. Similarly nuclear and endoplasmic \abbr{SPH} have high and almost equally distributed noise with most significant incident  reactions. For nuclear \abbr{SPH} the threshold was exceeded also by non-incident reactions 
incident with nuclear \abbr{CER}. For the membrane \abbr{SPH} species highly influential reaction is the \abbr{SPH} membrane diffusion. 
For the inner membrane \abbr{SPH} except the incident reactions  the high noise components steams from reactions connected with outer membrane \abbr{CER} 
(between outer membrane \abbr{SM} and \abbr{SPH}). The outer membrane \abbr{SPH} significant reactions include transportation reaction of \abbr{C1P} from Golgi apparatus to outer membrane and reactions 
connected with outer membrane \abbr{CER} and outer membrane \abbr{S1P}. For the lysosomal \abbr{SPH} except incident reactions the high noise component steams from reaction 
catalysed by \abbr{aSMase} in lysosome. For the cytoplasmic \abbr{SPH} except incident reactions the threshold was exceeded by the reaction 
catalysed by \abbr{AlkCDase} in Golgi apparatus. For Golgi apparatus \abbr{SPH} the noise was mainly decomposed by incident reactions.
\item The highest variability within \abbr{S1P} species has the mitochondrial \abbr{S1P}. It's all variance components are near to average noise and non of the reactions exceed the threshold of 110\%. Variance of all other \abbr{S1P} species steams principally from the incident reactions with one exception of cytoplasmic \abbr{S1P} which noise is generated mainly by lysosomal reaction 
catalysed by \abbr{aSMase}.
\item The \abbr{SM} species have the highest noise among all species and contrary to most other species the variance of \abbr{SM} species steams almost equally from all reactions.

\end{itemize}

  \subsubsection*{Sensitivity based parameters clustering  (homeostasis)}

The complex structure of our model makes it the ideal candidate to test the applicability of new method to detect the mutual relations between parameters via clustering of mutually compensative ones\cite{Wlodarczyk13}.
Usually parameters sets are not  pairwise independent and commonly a biochemical model is sensitive to the linear combination of parameters, what makes them not identifiable \cite{Brown03, Lipniacki04}.
\begin{figure}
\begin{center}
\textbf{Parameters clustering based on sensitivities – homeostasis}
\end{center} 
\includegraphics[width=\textwidth]{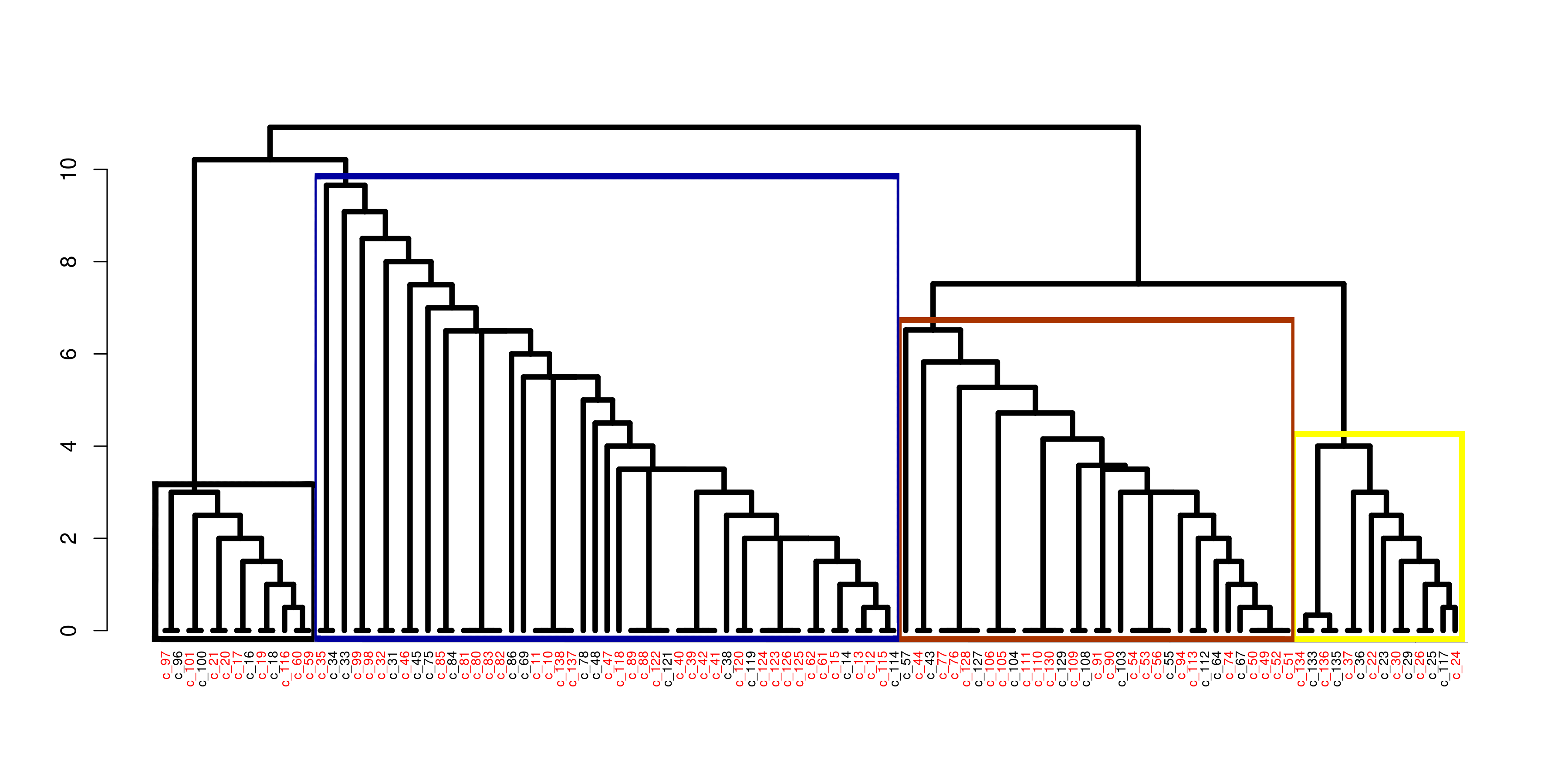}
\includegraphics[width=\textwidth]{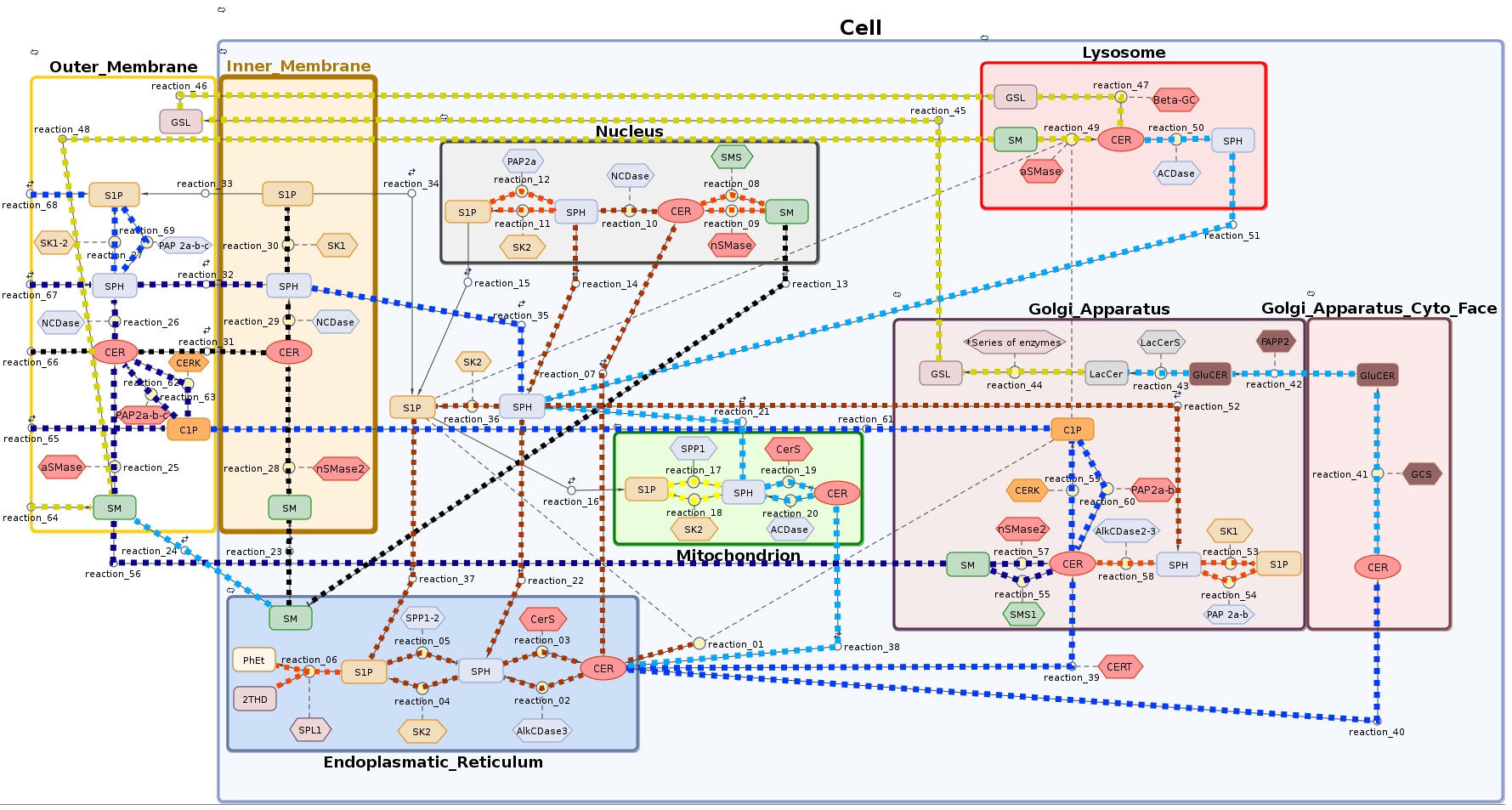}
\caption{(a) Dendrogram obtained by hierarchical clustering of parameters based on their functional redundancy.
Identifiability analysis yielded  37 non-identifiable parameters (marked  in red).
Correspondence between labels and names of the parameters is given  in supplementary materials.
 (b) Clusters of reactions induced by the hierarchical grouping. Colors of connections between species are compatible with colors of clusters in the dendrogram.
Higher color intensity within the cluster means greater level of redundancy between parameters of reaction and other parameters in the cluster. }\label{fig:PC_hom}
\end{figure}

Using sensitivity clustering of parameters (see Section Methods~\ref{sec:method})
we obtained the dendrogram with unequivocal division on four clusters (see Figure~\ref{fig:PC_hom}). We can interpret the obtained clusters as specific functional modules. Importantly, these results
are compatible with theoretical compartments recognized by Rao et al. ~\cite{rao}, who presented sphingolipid metabolic pathway as
a combination of following units: (i) C1 compartment represents the \textit{de novo} biosynthesis of \abbr{CER}, (ii) C2 compartment depicting the conversion of \abbr{CER} into complex sphingolipid like \abbr{SM} and \abbr{GSL}, (iii) C3 compartment represents hydrolysis of \abbr{SM} to \abbr{CER}, (iv) C4 compartment depicting conversion of \abbr{CER} into bioactive molecules such
as \abbr{C1P} and \abbr{S1P}.

\subsubsection*{Ceramide phosphorylation}

The structure of our clusters could be summarized as follows: brown cluster joins reactions parameters from endoplasmic reticulum, Golgi apparatus,  nucleus and cytoplasm. Strong redundancy between parameters reflects their functional correlation. This cluster is mostly related to the conversion of \abbr{SPH} to \abbr{S1P} and backwards. Beyoned sphingolipids phosphorylation and dephosphorylation it also contains reactions of sphingosine acetylation and reverse reaction – deacetylation of \abbr{CER}. To some extent it corresponds to the C4 compartment described by Rao et al.~\cite{rao}. However, our cluster does not include reactions taking place in cell membrane, which are classified to the blue and green cluster. Instead, we found reaction
depicting endogenous inflow of ceramide in endoplasmic reticulum (simulating the \textit{de novo} synthesis pathway) to be a part of brown cluster. It is worth mention that our analysis shows reactions in endoplasmic reticulum to form functional unity with nuclear reactions. What makes sense as, membranes of reticulum are structurally joined with the nuclear envelope.

\subsubsection*{Complex \abbr{SL} synthesis}

Blue cluster contains reactions parameters related to molecular composition of outer membrane. As cell membrane is the biggest reservoir of sphingolipids, especially complex sphingolipids like sphingomyelin and glycosphingolipids this cluster strongly affects general sphingolipid balance in the cell. Reactions from the pathway of \abbr{CER} production via \abbr{SM} hydrolysis, mentioned before, are localized in this cluster. This is confirmed by local sensitivity analysis, as we can observe strong influence of reaction
catalysed by \abbr{aSMase} in outer membrane to many modelled species.  As a consequence the pathways responsible for complex sphingolipids (\abbr{SM} and \abbr{GSL}) synthesis are localized in this cluster. These reactions the strongly affect stability of endoplasmic \abbr{CER} and subsequently cytoplasmatic \abbr{SPH}. Blue cluster is comparable to the C2 compartment as denoted in~\cite{rao} reflecting complex \abbr{SL} synthesis. However, extending the results from~\cite{rao}
we have shown that it forms a functional unity with reactions of outer cell membrane.  Interestingly results of our simulations are consistent with literature reports as \abbr{SM} metabolism at the plasma membrane is known to have strong implications for bioactive sphingolipids balance~\cite{milh10}.

\subsubsection*{Sphingolipds degradation}

Yellow cluster is related to the degradation of complex sphingolipids in acidic environment of lysosome. It includes the starting point of the \textit{salvage pathway} – \abbr{SM} transportation and degradation in lysosome and to some extent resembles the C3 compartment form~\cite{rao}
. According to the local sensitivity analysis two reactions from this cluster representing transportation of \abbr{SM} from outer membrane to lysosome and ceramide synthase from \abbr{SM} in lysosome 
may affect the concentration of different molecular species of the model like lysosomal and outer membrane \abbr{CER}, \abbr{SM} and \abbr{SPH}, endoplasmic \abbr{CER} or mitochondrial \abbr{S1P} and \abbr{SPH}. However, the strength of this influence is not very high. This finding can be explained by the relatively low activity of lysosomal degradation pathway in cells developing in favourable conditions. 
It should be mentioned that according to the clustering analysis, lysosome belongs to intersection of yellow and blue clusters what seems biologically correct as this organelle links pathways of complex \abbr{SL} synthesis and degradation. 

\subsubsection*{Inner membrane balance}

Black cluster reflects inner membrane molecular balance and contains reaction parameters which are not mutually related with other compartments, but have specific effect on behavior of other pathways. For instance this cluster contains reaction
catalysed by \abbr{nSMase} in inner membrane which, on the basis of local sensitivity analysis, appears to have slight impact on the stability of both \abbr{CER}, \abbr{SPH} and \abbr{S1P} in the whole model.

\subsection{Application of the model: case study of Alzheimer's disease}

Our model implements not only functional integration of experimental data but may be also used for computational verification of molecular changes known to cause various human diseases. In this study we applied our model to answer the question whether changes in enzymatic activity described by~\cite{rao} 
would lead to predicted cell behaviour typical for Alzheimer's disease (\abbr{AD}). In recent studies it became evident that sphingolipids play important roles in trafficking and metabolism of \abbr{AD} related proteins. Thus, they emerged to crucial molecules in etiology of \abbr{AD}~\cite{al3, al6}. 
This devastating neurodegenerative disorder is characterized by the accumulation of intraneuronal and extracellular protein aggregates and progressive synapse loss. Pathological hallmarks of \abbr{AD} are extracellular deposition of peptide termed $\beta$-amyliod (A$\beta$) and neurofibrillary tangles. Inability to catabolize aggregates of abnormally folded A$\beta$ leads to neuronal degeneration and subsequent decline in cognitive processes. On the level of sphingolipids metabolism most frequently reported are ceramide accumulation in endoplasmic reticulum and lysosome and sphingosine accumulation in cytoplasm accompanied by decreased level of cytoplasmatic \abbr{S1P} and \abbr{C1P}~\cite{al1, al2, al3, al4, al5}.

\subsubsection*{Computational simulation of Alzheimer's disease}
   
Aiming to simulate cell response to metabolic disturbances of \abbr{SL} pathway described in \cite{rao} we have changed values of chosen reaction parameters for cell homoeostasis. For details see supplementary Table~3. 
We have modified parameters corresponding to the ceramidase (\abbr{CDase}) activity as well as parameters corresponding to sphingosine kinase (\abbr{SK}) and ceramide kinase (\abbr{CERK}) dynamics. Moreover due to down-regulation of \abbr{CERT} expression we have inhibited transportation of \abbr{CER} to Golgi apparatus. On the other hand ceramide  \textit{de novo} synthesis reflected by the inflow reaction
of \abbr{CER} in endoplasmic reticulum was up-regulated. To test the system response in a \abbr{AD} scenario we simulated the time evolution of species concentrations.

 \begin{figure}[!ht]
   \begin{center}
	\textbf{Trajectories of species concentration - \abbr{AD} scenario}\\
	 \end{center}
    \includegraphics[width=0.5\textwidth]{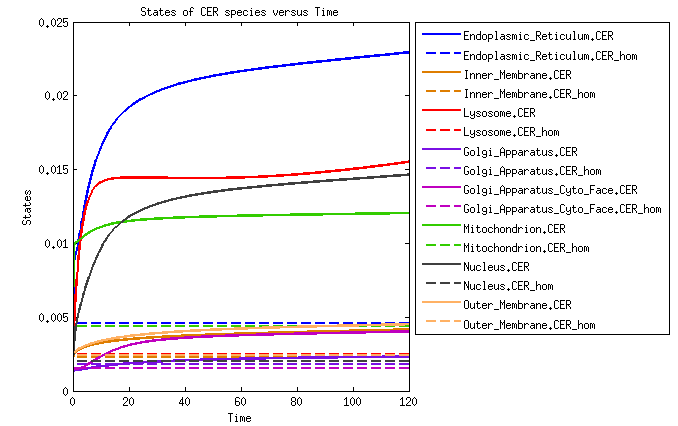}
    \includegraphics[width=0.5\textwidth]{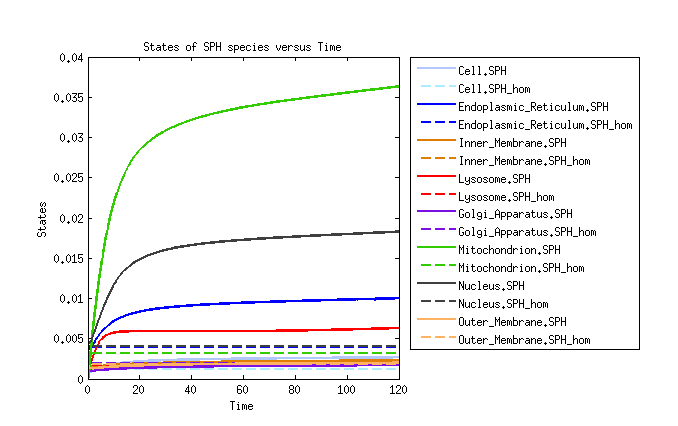}\\
    \includegraphics[width=0.5\textwidth]{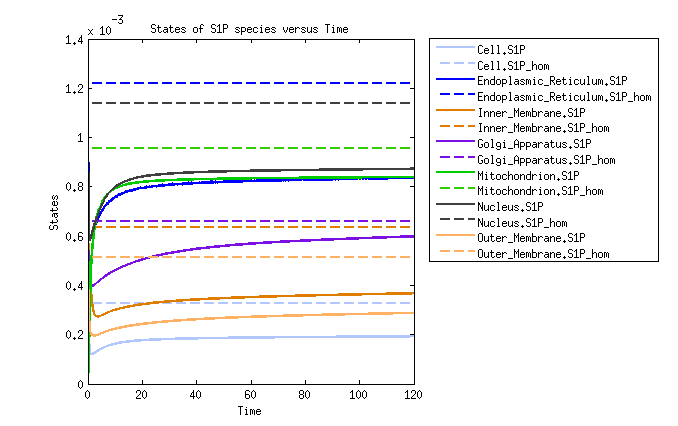}
    \includegraphics[width=0.5\textwidth]{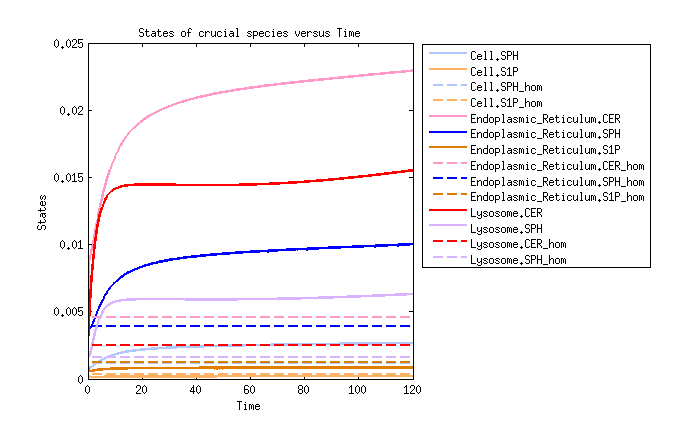}
   \caption{Time evolution of molar concentration for the following groups of species  (dashed lines correspond to homeostasis scenario, solid lines to AD scenario):  (a)  ceramides species; (b) sphingosine species; (c)~sphingosine-1-phosphate species; (d) species functionally related to AD.}\label{fig:traj_AD}
 \end{figure}

Preliminary simulations have shown, that when changes were limited to those described by Rao et al.~\cite{rao} some modelled species quickly diverged to infinity. Namely, we observed unexpected rapid  cumulation of \abbr{SM} in many cellular compartments (i.e. lysosome, outer membrane and endoplasmic reticulum). Another unforeseen system behaviour was increased rate of \abbr{CER} to \abbr{GSL} conversion in Golgi Apparatus, followed by accumulation of \abbr{GSL}. Since such events do not occur in \abbr{AD} cells, we suggested that, initially introduced modifications should be accompanied by: (i)reduced transportation of \abbr{CER} to Golgi apparatus via \abbr{CERT} independent pathway, (ii) increased activity of sphingomyelinases (\abbr{SMase}). We also introduced some minor changes in \abbr{SM} transportation between compartments. Importantly our predictions were confirmed by the literature, namely impaired \abbr{SM} metabolism is known to be linked with \abbr{AD}~\cite{jana10, lee14}. These findings emphasize the predictive value of our model. 

After these biologically justified modifications all our results were coherent with experimental data, namely we observed cumulation of ceramides in cellular compartments, especially in endoplasmic reticulum (\abbr{ER}) and lysosome, in comparison to the homoeostasis level~\cite{al6, rao}.   

The model output for sphingosine species concentration showed the imitate drop of \abbr{SPH} species due to \abbr{CDase} down regulation, but than cumulation of \abbr{SPH} species in all compartments as a consequence of the increase of \abbr{CER} species concentration. We also recorded the decrease of \abbr{S1P} species concentration in the \abbr{AD} scenario (see Figure \ref{fig:traj_AD}).

\subsubsection*{Local sensitivity analysis for the \abbr{AD} scenario vs homeostasis}

Application of the \abbr{AD} scenario 
yielded slight changes in parameters local sensitivities.  
\begin{itemize}
\item For the \abbr{CER} species, in contrary to homoeostasis, the most sensitive become ceramides in nucleus and endoplasmic reticulum, that are sensitive basically to endogenous \abbr{CER} in endoplasmic reticulum and exogenous \abbr{C1P}, \abbr{CER}, \abbr{SPH}, \abbr{S1P} in outer membrane inflow parameters as well as \abbr{nSMase} reaction rate in outer membrane.
\item The \abbr{S1P} in cytosol becomes sensitive to the \abbr{SK2} in cytosol
, analogously \abbr{S1P} in inner membrane becomes sensitive to \abbr{SK1} in inner membrane and \abbr{S1P} in outer membrane is more sensitive to inflow parameter
of exogenous \abbr{S1P} in outer membrane. However the mitochondrial \abbr{S1P} becomes invariant to parameters changes.
\item \abbr{SPH} species remains largely unchanged with most sensitive mitochondrial \abbr{S1P}.
\item Similarly \abbr{SM} species show an unchanged sensitivity with the dominant species \abbr{SM} in outer membrane most sensitive.
\end{itemize}

\subsubsection*{Parameters clustering in the \abbr{AD} scenario vs homeostasis}
 
Clustering analysis of \abbr{AD} model resulted in new parameters dendrogram with only two clusters in comparison to four clusters obtained in homoeostasis (see Figure \ref{fig:PC_AD} and compare to Figure~\ref{fig:PC_hom}). Clusters distinguished in \abbr{AD} simulation can be described as follows.

\begin{figure}[!ht]
\begin{center}
\textbf{Parameters clustering based on sensitivities – \abbr{AD} scenario}\\
\end{center} 
\includegraphics[width=\textwidth]{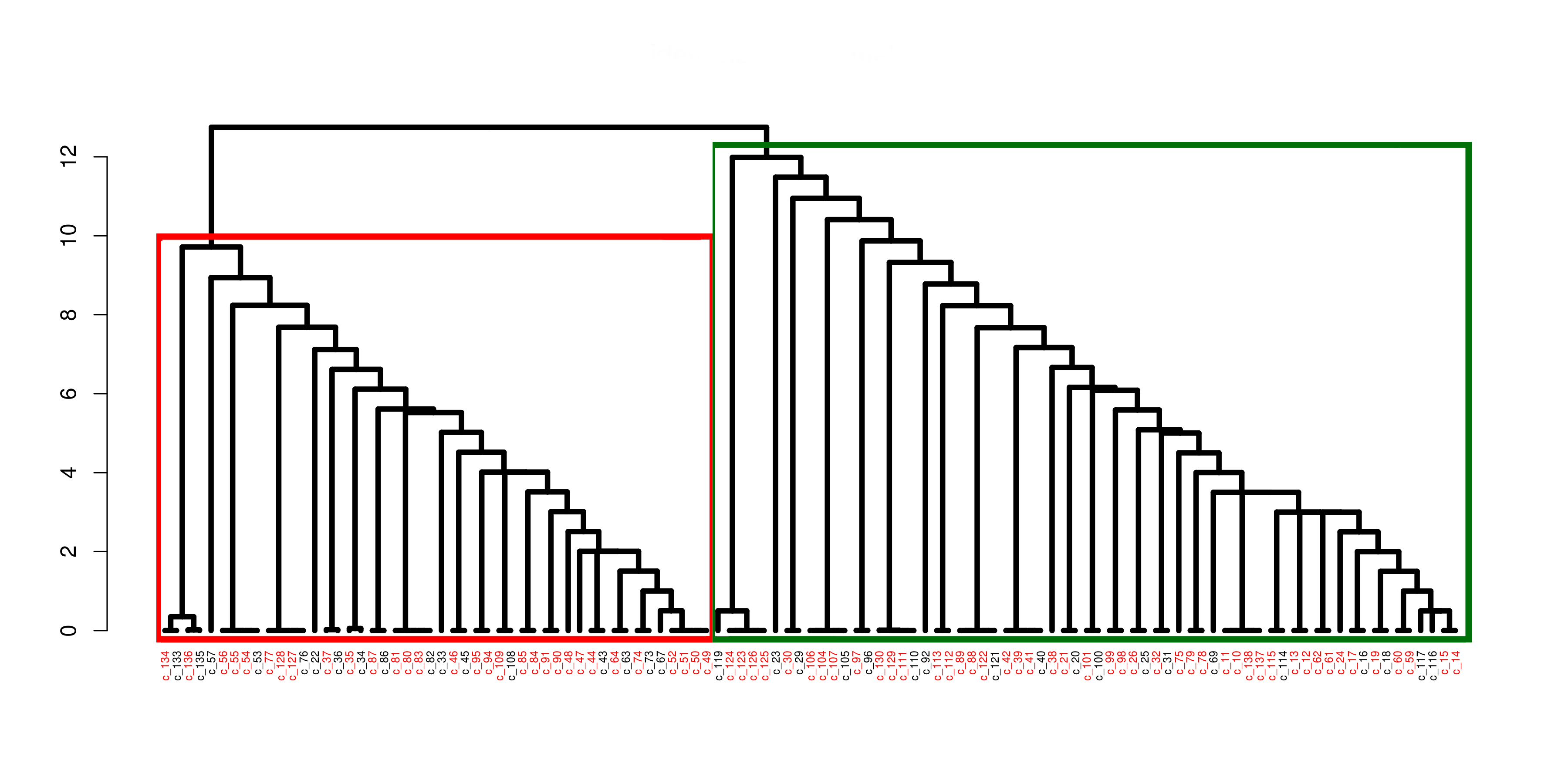}
\includegraphics[width=\textwidth]{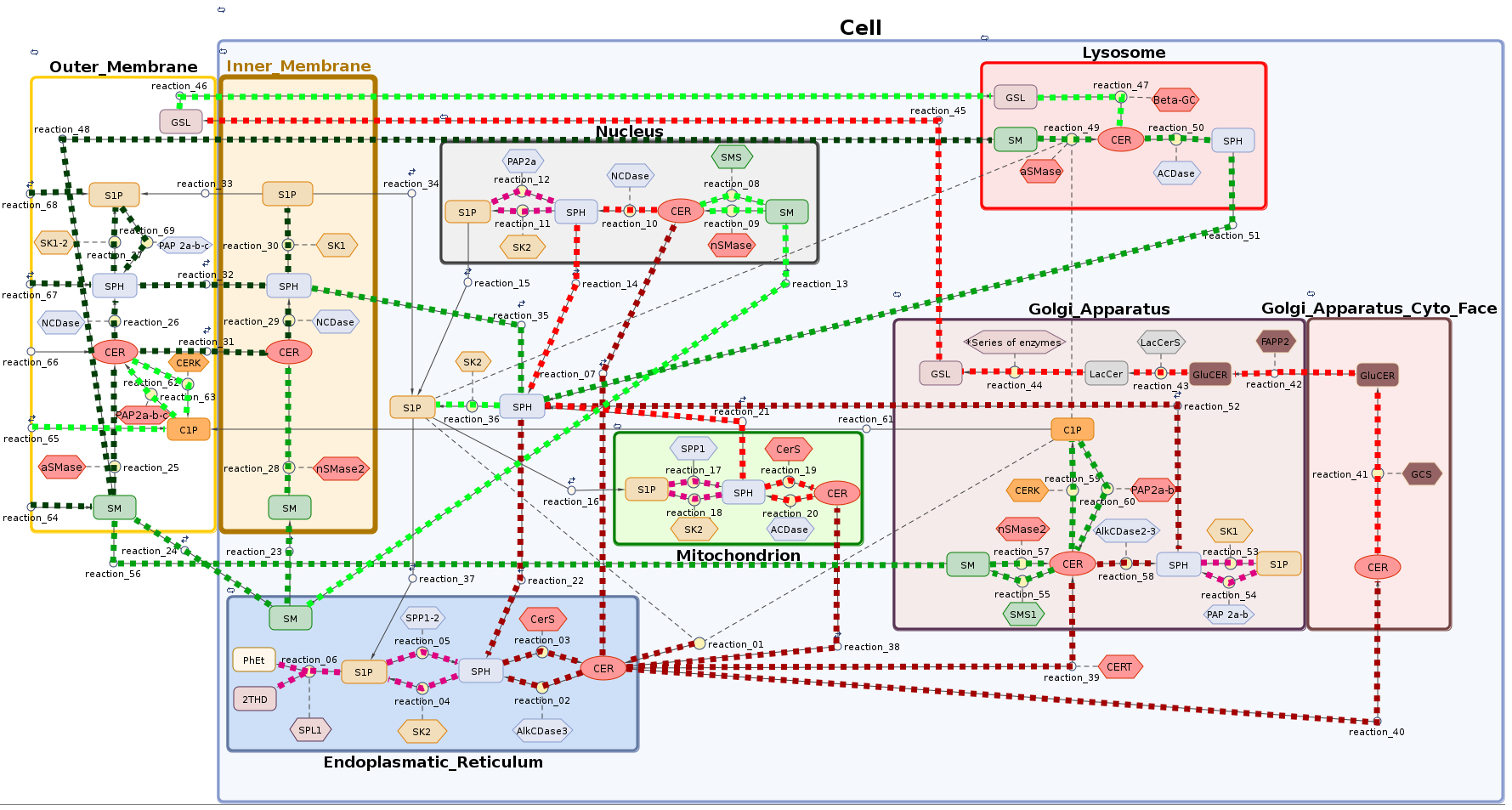}
\caption{(a) Dendrogram obtained for AD scenario by hierarchical clustering of parameters based on their functional redundancy. Model contains 36 non-identifiable parameters.  (b) Clusters of reactions induced by the hierarchical grouping. }\label{fig:PC_AD}
\end{figure}

\subsubsection*{Complex SL and \abbr{GSL} metabolom}

Green cluster is grouping parameters strongly associated with complex sphingolipids and glycosphingolipids metabolism. It comes out that changes of \abbr{SM} balance are of great importance for cellular metabolism in \abbr{AD}. This cluster includes formation of ceremide on the pathway of \abbr{SM} hydrolysis, catabolism of \abbr{SM} and \abbr{GSL} in lyzosom  (\textit{salvage pathway}). Similarly to results obtained in homoeostasis our model confirms that hydrolysis of \abbr{SM} in cell membrane and lysosome strongly influences the level of cytoplasmic \abbr{SPH}. This cluster also includes parameters related to the synthesis of \abbr{GSL} and \abbr{SM} in Golgi apparatus.   To conclude this cluster can be regard as a combination of green, yellow and a part of the blue homeostatic cluster its formation can be explained by the growth of the importance of \abbr{SM} transportation and degradation during neurodegeneration. 

\subsubsection*{Ceramide synthesis and accumulation}

Red cluster includes reactions mostly affected by the inflow of the ceramide from \textit{de novo} synthesis pathway. According to lit.\cite{pas05} in \abbr{AD} endoplasmic accumulation of ceramide from this source is important step in disease development. Clusters analysis confirmed strong correlation between \textit{de novo} synthesis and concentration of \abbr{CER} in endoplasmic reticulum and subsequently in mitochondrion, nucleus and also Golgi apparatus.

\section{Conclusions}
In the present paper an original model for  sphingolipid  metabolism in non-specified human tissue was proposed. 
To our best knowledge, it is  the most comprehensive model so far and also the first one, that explicitly comprises the compartmentalization.
We managed to keep the balance between the complexity and biological soundness of the model and its computational tractability.

Our results demonstrate that the model is an excellent tool for predicting SL pathway reaction to perturbation in activity of particular enzymes, 
as well as the up- or down-regulation of modeled species.  Thus the model may perfectly be used to simulate molecular behavior in various scenarios as it was shown in the case study of Altzheimer's disease.

Moreover, the implementation  of semi-independent compartments (and the transportation reactions for the molecular species flow between them) 
allows for more subtle manipulations of the reaction parameters for  specific organelle. Finally our model enables not only the integration but also validation of experimental data by 
verifying  their cross-compliance in a complex network of interactions.

Last but not least,  the computational validation of the model was performed by means of recently proposed sophisticated approaches~\cite{komor2013, Wlodarczyk13}. 
Mathematically elegant methods of variance decomposition and sensitivity clustering of parameters 
revealed non-trivial biological outcomes.
More importantly, at the same time the application of above mentioned approaches  to our model constituted the perfect validation of their usefulness in realistic size problems.

\section{Methods}\label{sec:method}

All molecular reactions within a system of interacting species $S_1\ldots S_N$ may be presented in the following manner:
\begin{equation*}
        R_j \colon \qquad \underline{\nu}_{1j} S_1 + \dots + \underline{\nu}_{Nj} S_N \stackrel{k_j}        		\longrightarrow \overline{\nu}_{1j} S_1 + \dots + \overline{\nu}_{Nj} S_N,
\end{equation*}
where $\underline{\nu}_{nj}$ and $\overline{\nu}_{nj}$ denote amounts of molecules of $n$-th
species that are respectively substrate and product of this reaction and the coefficient $k_j$ denotes reaction rate (speed) of the reaction.

\subsubsection*{The Mass Action Law kinetics.}
In case of non-enzymatic or transpiration kinetics  we used the Mass Action Law (\abbr{MAL}) principle.  The time derivative of each species concentration is the sum of in- and out-fluxes of all neighbouring reactions. Here the one reaction flux is equal $k_j \left[S_1\right]^{\underline{\nu}_{1j}} \cdot \dots \cdot \left[S_N\right]^{\underline{\nu}_{Nj}}$. Hence the Ordinary Differential Equations derived from the \abbr{MAL} can be expressed in the form:
\begin{equation*}
		\frac{d[S_n]}{dt} = \sum_{j=1}^{R} s_{nj} k_j \left[S_1\right]^{\underline{\nu}_{1j}} \cdot 				\dots \cdot \left[S_N\right]^{\underline{\nu}_{Nj}} \qquad n=1 \dots N
\end{equation*}
where  $s_{nj} = \overline{\nu}_{nj} - \underline{\nu}_{nj}$  denotes a stoichiometric coefficient of $n$-th species in $j$-th reaction and $[S_n]$ denotes the concentration of $n$-th species.

\subsubsection*{The Michaelis-Menten kinetics.}

Most of the reactions depicted in the diagram \ref{fig:diagram} are enzymatic reactions. For this kind of  reactions we used Michaelis-Menten model (\abbr{MM}) and simplified kinetics derived by the \abbr{MM} model: 
%
	\begin{equation*}
		\frac{d[P]}{dt}=\frac{V_{max}[S]}{K_m+[S]} ,
	\end{equation*}
where $P$ denotes reaction product, $S$ denotes reaction substrate and 	$V_{max}$, $K_m$ are constant reaction parameters.

\subsubsection*{Ordinary Diferential Equations.}

Equivalently the \abbr{ODEs} can be expressed in the matrix form:
	\begin{equation*}
        \frac{d\xVec{S}(t)}{dt} = M
        \xVec{v}(\xVec{S}(t)),
	\end{equation*}
where the system state is represented by the time dependent state vector $\xVec{S}(t)$ of species concentration, $M$ denotes the stoichiometry matrix and $\xVec{v}(\xVec{S}(t))$ denotes a vector of reaction fluxes (in our model according to \abbr{MAL} or \abbr{MM} kinetics including inhibition rates)\cite{chara}.

\subsection{Local Sensitivity Analysis}

Local sensitivity analysis shows how the uncertainty of parameters of the model can influence the model output. Sensitivity may be measured by monitoring changes in the output by e.g. partial derivatives of the modelled species to the single parameters. This appears a logical approach as any change observed in the output will unambiguously be due to the single variable changed. To compare the sensitivity of the model to the single parameters one construct the sensitivity indices by time integration of partial derivatives:
$$
s_{n,i}=\int_0^T\left\vert\frac{\partial S_n(t)}{\partial \theta_i}\right\vert_{\theta=\theta_0}dt
$$
where $S_n$ are different species concentrations, $\theta$ is the vector of parameters and $\theta_0$ is some fixed  point in parameters space. 

\subsection{Variance Decomposition}

The  deterministic approach that represents the mean behaviour of the system  can also be generalized to a stochastic mode by meaning of Stochastic Differential Equations (\abbr{SDE}), both of which can be represented in a discrete Markov Chain or a continuous Markov Process. Below we sketch the method of variance decomposition as presented in \cite{komor2013}.

\subsubsection*{Stochastic Differential Equations.}

Modelling the system behaviour in a stochastic manner means the examination  not only the evolution of the average system state that represents one of possible trajectories, but examination of the evolution of the probability distribution over all possible system states.

The most popular approach to describe discrete stochastic model of biochemical pathway is Chemical Master Equation (Chapman-Kolmogorov equation of Markov chain modelling the evolution of the system):
	\begin{equation*}
        \frac{p P(\xVec{x},t)}{dt} =
		\sum_{j}a_j(\xVec{x}-\xVec{m}_j)P(\xVec{x}-\xVec{m}_j,t) -
		\sum_{j}a_j(\xVec{x})P(\xVec{x},t),
    \end{equation*}
where the system state is denoted by the vector $\xVec{X}(t)\in\mathbb{N}^{N}$ of numbers of molecules each row for one of $N$ reacting species, $\xVec{m}_j$ denotes the $j$-th column of stoichiometry matrix $M=(\xVec{m}_1,\dots ,\xVec{m}_{R})$ and $P(\xVec{x}, t)$ denotes the time- and state-dependent distribution of system being in state $\xVec{X}(t) = \xVec{x}$ and finally $a_j(\xVec{X}(t))$ denotes the propensity function associated with the $j$-th reaction\cite{chara}.

One of the possible simplification of the above equation is Linear Noise Approximation, where the dynamic is modelled with Poisson process:     
	\begin{equation*}
		\xVec{X}(t) = \xVec{X}(0) + \sum\limits_{j = 1}^R {{\xVec{m}_j}{N_j}\left( {\int\limits_0^t {{f_j}(\xVec{X}(\tau),\tau)d\tau} } \right)}
	\end{equation*}
where $N_j(\xVec{X}(t),t)$ denotes Poisson process dependent on time and a system state $\xVec{X}(t)$, corresponding to occurrence of $j$-th reaction. The probability that $j$-th reaction occur during the time interval $[t; t+dt)$ equals $f_j(x, t)dt$, where the $f_j(x, t)$ is called the transition rate. 
    
Although accurate discrete models describe the exact  evolution of probability distribution of the system with the assumption that in one time point at most one reaction can occurre, they are computational not efficient, as simulations require significant resources. Consequently it is more efficient to transit from discrete to continuous process. Starting from deterministic approximation:
	\begin{equation*}
		\Phi(t) = \Phi(0) + \sum\limits_{j = 1}^R {{m_j} {\int\limits_0^t {{f_j}(\Phi(s),s)ds} }}
	\end{equation*}
where $\Phi(t)$ is the mean system state being the solution of the ODEs one can describe the system state evolution by dividing it into deterministic and stochastic part:
	\begin{equation*}
		x(t) = \xi(t)+ \Phi(t)
	\end{equation*}
where $\Phi(t)$ is the deterministic part and $\xi(t)$ is the Winer process describing stochastic noise of a system state\cite{komor2009}.
The next step of stochastic noise decomposition is to divided noise linearly into noise steaming from separate reactions. 
The fact, that  the total variance:
$$
\Sigma(t)=\langle (x(t) - \langle x(t)\rangle) (x(t) - \langle x(t)\rangle)^T  \rangle
$$
is described by the differential equation
\begin{equation}\label{dVar}
\frac{d\Sigma}{dt}={{A}}(t)\Sigma+ \Sigma{{A}}(t)^T + {{D}}(t),
\end{equation}
where 
$$
\left\{ A(\Phi,t) \right\}_{ik}=\sum_{j=1}^r m_{ij}
\frac{\partial f_j(\Phi,t)}{\partial \Phi_k}
$$ and $D(t)$ denotes diffusion matrix,
 can be represented as the sum of individual contributions,
\begin{equation}
\Sigma(t)=\Sigma^{(1)}(t)+\ . . .\  +\Sigma^{(r)}(t).
\end{equation}
results directly from the decomposition of the diffusion matrix  ${{D}}(t)=\sum_{j=1}^{r} {{D}}^{(j)}(t)$ and
the linearity of the equation for $\Sigma(t)$.
\cite{komor2013}
By decomposing variance into  the components from individual reactions, we are able to determine the variability that the model has from each reaction, and therefore we are able to assess and weigh the uncertainty of the model in division into single reactions.

\subsection{Parameters clustering}

In  \cite{Wlodarczyk13} the concept of {\em functional redundancy} has been proposed and used in hierarchical clustering algorithm as dissimilarity measure.
Let us define the model in Bayesian approach by the distribution of data  $(X \in \mathbb{R}^{k})$ given parameters 
$(\theta \in \mathbb{R}^{l})$ as $P(X | \theta)$, together with \textsl{a priori} distribution $P(\theta)$.
Assume that $\theta = (\theta_{A}, \theta_{B})$ corresponds to division of parameters on two independent sets then \cite{Ludtke08}:
\begin{align}
H(X) = I(X, \theta_{A}) + I(X, \theta_{B}) + I(\theta_{A}, \theta_{B}| X) + H(X|\theta),
\end{align}
where $H$ denotes here the entropy and $I$ is the mutual information between random variables. Here
$I(\theta_{A}, \theta_{B}| X)$ measures this part of entropy which is shared by both sets of parameters 
and it is equivalent to redundant knowledge of the model which is owned by $\theta_{A}$ and $\theta_{B}$. 

Computation of $I(\theta_{A}, \theta_{B}| X)$ requires calculation integral over all possible outcomes of the model what is highly inefficient, 
hence, this notion has been replaced with local redundancy measure 
which substitute assumption of knowledge about the model $X$ with information about initial parameters $\theta^{*}$.
Thus, functional redundancy is equal to $I(\theta_{A}, \theta_{B}|\theta^{*})$ and is calculated according to the formul \cite{Johnson92}:
\begin{align}\label{eq:fr_frComp}
I(\theta_{A}, \theta_{B}|\theta^{*}) = -\frac{1}{2}\sum_{i = 1}^{\text{min}(|\theta_{A}|, |\theta_{B}|)} \text{log}(1 - \rho_{j}^{2}),
\end{align} 
where $\rho_{i}$ stands for the canonical correlation obtained from the Fisher Information Matrix of  $\theta^{*}$ ($FIM(\theta^{*})$).

Moreover, to indicate {\em non-identifiable} parameters authors define term of $(\delta, \zeta)$ - identifiability
using the idea of functional redundancy \cite{Wlodarczyk13}.
In this terminology $\theta_{i}$ is $(\delta, \zeta)$ - identifiable if $FIM_{ii}(\theta) > \zeta$
and $\rho(\theta_{i}, \theta_{-i}) < 1 - \delta$, 
where $\theta_{-i}$ represents all parameters except $\theta_{i}$. 

Using functional redundancy we can cluster parameters according to hierarchical algorithm 
(i.e. in every turn of the loop we merge two sets of the parameters with the biggest redundancy measure 
and remove all nonindentifiable parameters from further analysis) and visualize it on a dendrogram.

\section*{Authors contributions}
  WW designed the biochemical model, AC and KN implemented it and carried out the computational experiments. AG inspired the research  and supervised the project. All authors contributed to the writing of this manuscript and have read and approved the final manuscript.
  
 \section*{ Competing interests}
The authors declare that they have no competing interests.

\section*{Acknowledgements}
  \ifthenelse{\boolean{publ}}{\small}{}
All authors thank prof. Bogdan  Lesyng and prof. Robert  Strosznajder (Mossakowski Medical Center, Polish Academy of Sciences) for valuable discussions and for inspiring this research and, above all -- Micha{\l} Komorowski
for acquainting us with the model analysis methods.
This work was partially supported by Polish National Science Center grant $\text{n}^\text{o}$ 2011/01/B/NZ2/00864, Biocentrum-Ochota project (POIG 02.03.00-00-003/09) and EU project  POKL.04.01.01-00-051/10-00.

\newpage


\printglossaries
 

{\ifthenelse{\boolean{publ}}{\footnotesize}{\small}
\newpage

 \bibliographystyle{sfingo}  
  \bibliography{sfingo} }     


\ifthenelse{\boolean{publ}}{\end{multicols}}{}



\begin{stare}
\section*{Figures}
  \subsection*{Figure 1 - The ceramides metabolism diagram}
      A short description of the figure content
      should go here.


  \subsection*{Figure 2 - The local sensitivity analysis of ceramides species}
      Figure legend text.


  \subsection*{Figure 3 - The variance decomposition for ceramides species}
      A short description of the figure content
      should go here.
      

  \subsection*{Figure 4 - Trajectories of species concentration - AD scenario}
      Figure legend text.


  \subsection*{Figure 5 - Parameters clustering based on sensitivities – homeostasis}
      Figure legend text.


  \subsection*{Figure 6 - Parameters clustering based on sensitivities – AD scenario}
      A short description of the figure content
      should go here.
      


\section*{Tables}

    
%
%
%

%

\subsection{Table 1 - Parameters of the ceramides metabolism model}
\begin{sidewaystable}[!ht]
 	\begin{center}
	\begin{tiny}
 	\begin{tabular}{|
 p{0.01\paperwidth}p{0.08\paperwidth}p{0.15\paperwidth}p{0.04\paperwidth}p{0.09\paperwidth} p{0.16\paperwidth}p{0.035\paperwidth}p{0.045\paperwidth}p{0.04\paperwidth}|}
\hline
& & & & & & & &\\
\parbox[t]{0.5pt}{\textbf{RX}\\\textbf{no.}}&\textbf{Compartment}&\textbf{Reaction Product}&\textbf{Kinetics}&\textbf{Reaction Flux}&\textbf{Enzyme}&\parbox[t]{0.5pt}{$\mathbf{Km}$\\$\mathbf{(\mu \abbr{M})}$}&\parbox[t]{0.5pt}{$\mathbf{Vm}$\\$\mathbf{(\mu M/min)}$}&\parbox[t]{0.5pt}{$\mathbf{k_j}$\\$\mathbf{(1/min)}$}\\
\hline
& & & & & & & &\\
1&\parbox[t]{1in}{Outer \\Membrane}& \parbox[t]{2in}{Sphingosine 1\nobreakdash-phosphate \\(OM.S1P)}&MM& \parbox[b]{1in}{$\frac{Vm1*OM.SPH}{Km1+OM.SPH}$}&\parbox[t]{2in}{ Sphingosine Kinase~1\nobreakdash-2\\(SK1\nobreakdash-2)}&3,4&0,4&-\\
2&\parbox[t]{1in}{Outer \\Membrane}&Sphingosine (OM.SPH)&MM& \parbox[b]{1in}{$\frac{Vm2*OM.CER/}{(Km2+OM.CER)}$}&nCeramidase (nCDase)&1&1&-\\
3&\parbox[t]{1in}{Outer \\Membrane}&Ceramide (OM.CER)&MM&\parbox[t]{0.5pt}{Vm3*OM.SM/\\(Km3+OM.SM)}&Neutral sphingomyelinase (\abbr{nSMase})&0,0002&0,0583&-\\
4&Inner Membrane&Sphingosine 1-phosphate (IM.S1P)&MM&\parbox[t]{0.5pt}{Vm4*IM.SPH/\\(Km4+IM.SPH)}&Sphingosine Kinase1 (\abbr{SK1})&5,06&0,94&-\\
5&Inner Membrane&Sphingosine (IM.SPH)&MM&Vm5*IM.CER/(Km5+IM.CER)&nCeramidase (nCDase)&1&1&-\\
6&Inner Membrane&Ceramide (IM.CER)&MM&Vm6*IM.SM/(Km6+IM.SM)&Neutral sphingomyelinase (\abbr{nSMase})&0,0002&0,0583&-\\
7&Cell&Glycosphingolipid (IM.GSL)&MAL&k7*GA.GSL&-&-&-&1\\
8&Cell&Glycosphingolipid (L.GSL)&MAL&k8*IM.GSL&-&-&-&1\\
9&Lysosome&Sphingomyelin (L.SM)&MAL&k9*IM.SM&-&-&-&1\\
10&Lysosome&Ceramide (L.CER)&MM&Vm10*L.SM/(Km10+L.SM)&Acid sphingomyelinase (\abbr{aSMase})&0,0002&0,0583&-\\
11&Lysosome&Ceramide (L.CER)&MM&Vm11*L.GSL/(Km11+L.GSL)&Acid beta-glucosidase (Gcase)&30&10&-\\
12&Lysosome&Sphingosine (L.SPH)&MM&Vm12*L.CER/(Km12+L.CER)&Acidic Ceramidase (\abbr{ACDase})&149&2,26&-\\
13&Golgi Apparatus&Glucosylceramide (GA.GluCER)&MM&Vm13*GAOL.GluCER/(Km13+GAOL.GluCER)&(\abbr{FAPP2})&10&10&-\\
14&Golgi Apparatus&GA.LacCer&MM&Vm14*GA.GluCER/(Km14+GA.GluCER)&(LacCerSase)&1&1&-\\
15&Golgi Apparatus&Glycosphingolipid (GA.GSL)&MM&Vm15*LacCer/(Km15+LacCer)&(Series of enzymes)&1&1&-\\
16&Cell&Sphingomyelin (IM.SM)&MAL&k16*GA.SM&-&-&-&1\\
17&Golgi Apparatus&Sphingomyelin (GA.SM)&MM&Vm17*GA.CER/(Km17+GA.CER)&Sphingomyelin synthase 1 (\abbr{SMS1})&7,49&27&-\\
18&Golgi Apparatus&Ceramide (GA.CER)&MM&Vm18*GA.SM/(Km18+GA.SM)&Neutral sphingomyelinase2-3 (nSMase2-3)&10&10&-\\
19&Golgi Apparatus&Sphingosine (GA.SPH)&MM&Vm19*GA.CER/(Km19+GA.CER)&Alkaline Ceramidase2-3 (AlkCDase2-3)&10&10&-\\
20&Cell&Sphingosine (ER.SPH)&MAL&k20*GA.SPH&-&-&-&1\\
21&Golgi Apparatus&Sphingosine 1-phosphate (GA.S1P)&MM&Vm21*GA.SPH/(Km21+GA.SPH)&Sphingosine Kinase1 (\abbr{SK1})&1&1&-\\
22&Endoplasmatic Reticulum&Ceramide (ER.CER)&MAL&k22&-&-&-&10\\
23&Golgi Apparatus Outer Layer&Glucosylceramide (GAOL.GluCER)&MM&Vm23*GAOL.CER/(Km23+GAOL.CER)&Glucosylceramide synthase (\abbr{GCS})&10&10&-\\
24&Cell&Ceramide (GA.CER)&MAL&k24*ER.CER&-&-&-&1\\
25&Cell&Ceramide (ER.CER)&MAL&k25*ER.CER&-&-&-&1\\
26&Endoplasmatic Reticulum&Ceramide (ER.CER)&MM&Vm26*ER.SPH/(Km26+ER.SPH)&Neutral sphingomyelinase (\abbr{nSMase})&1&1&-\\
27&Endoplasmatic Reticulum&Sphingosine (ER.SPH)&MM&Vm27*ER.CER/(Km27+ER.CER)&Alkaline Ceramidase3 (\abbr{AlkCDase3})&1&1&-\\
28&Endoplasmatic Reticulum&Sphingosine (ER.SPH)&MM&Vm28*ER.S1P/(Km28+ER.S1P)&Sphingosine 1-phosphatase1-2 (SPP1-2)&38,5&36,4&-\\
29&Endoplasmatic Reticulum&Sphingosine 1-phosphate (ER.S1P)&MM&Vm29*ER.SPH/(Km29+ER.SPH)&Sphingosine Kinase2 (\abbr{SK2})&10&10&-\\
30&Endoplasmatic Reticulum&PhEt+2THD&MM&Vm30*ER.S1P/(Km30+ER.S1P)&Sphingosine 1-phosphate lyase1 (S1PL1)&35&4,5&-\\
31&Membrane&Sphingomyelin (OM.SM)&MAL&k31*IM.SM&-&-&-&1\\
32&Membrane&Sphingosine (IM.SPH)&MAL&k32*OM.SPH&-&-&-&1\\
33&Membrane&Sphingosine 1-phosphate (OM.S1P)&MAL&k33*IM.S1P&-&-&-&1\\
34&Outer Membrane&-&MAL&k34*OM.S1P&-&-&-&1\\
35&Cell&Sphingosine 1-phosphate (ER.S1P)&MAL&k35*IM.S1P&-&-&-&1\\
36&Cell&Sphingosine (IM.SPH)&MAL&k36*IM.SPH&-&-&-&1\\
37&Cell&Sphingosine (ER.SPH)&MAL&k37*L.SPH&-&-&-&1\\
38&Cell&Sphingosine (IM.SPH)&MAL&k38*L.SPH&-&-&-&1\\
\hline
 	\end{tabular}
 	\end{tiny}
 	\end{center}
\caption{ }\label{tab:reac_par}
 \end{sidewaystable} 

  \subsection*{Table 2 - Supplementary: Initial values}

\begin{table}

\caption{ }\label{tab:init}
\end{table}

  \subsection*{Table 3 - Supplementary: Parameters in AD}

\begin{table}

\caption{ }\label{tab:par_AD}
\end{table}


\section*{Additional Files}
  \subsection*{Additional file 1 --- Sample additional file title}
    Additional file descriptions text (including details of how to
    view the file, if it is in a non-standard format or the file extension).  This might
    refer to a multi-page table or a figure.

  \subsection*{Additional file 2 --- Sample additional file title}
    Additional file descriptions text.
\end{stare}

\end{bmcformat}
\end{document}